\documentclass[prl,twocolumn]{revtex4}
\usepackage{graphicx,amssymb}
\usepackage{graphicx}
\usepackage{rotating}
\usepackage{dcolumn}
\usepackage{amsfonts}
\usepackage{amssymb}
\usepackage{amsmath}
\usepackage{latexsym}
\usepackage{epsfig}
\usepackage{dsfont,eucal,mathrsfs}
\usepackage{bm}

\usepackage{color}

\renewcommand{\mathbf}{\boldsymbol}
\renewcommand{\mathcal}{\mathscr}

\begin{document}

\title{Connectivity driven Coherence in Complex Networks}
\author{Tiago Pereira$^{1}$}
%\email{tiago.pereira@imperial.co.uk}
\author{Deniz Eroglu$^{2}$,  G. Baris Bagci$^{2}$, Ugur
Tirnakli$^{2}$}\author{Henrik Jeldtoft Jensen$^{1}$}
%\email{h.jensen@imperial.ac.uk}
\affiliation{$^1$Complexity\&Networks Group and Department of Mathematics,
 Imperial College London, London, UK}
\affiliation{$^2$ 
Department of Physics, Faculty of Science,
Ege University, 35100 Izmir, Turkey}

\begin{abstract}
We study the
emergence of coherence in complex networks of mutually coupled non-identical elements. 
We uncover the precise dependence of the dynamical coherence on the network connectivity, 
on the isolated dynamics of the elements and the coupling function.  These findings predict that 
in random graphs, the enhancement of coherence 
is proportional to the mean degree. In locally 
connected networks, coherence is no longer controlled by the mean degree, but rather on how the mean degree scales with the network size. In these networks, even when the coherence is absent, { adding a fraction $s$ of random connections leads to an enhancement of coherence proportional to $s$}. Our results provide a way to control
the emergent properties by the manipulation of the dynamics of the elements and the network connectivity.
\end{abstract}

\maketitle
Among the large variety of dynamical phenomena observed in complex networks, 
collective behavior is ubiquitous and has proven to be essential to the network function \cite{Fries,PS,Tiago,Attention,Singer,Ep,Gorka,Kurths}.   During the last decades, our 
understanding of collective behavior of complex networks has increased significantly. 
Most research focuses on synchronization of diffusively coupled of periodic elements  with distinct  frequencies
\cite{Strogatz1,Strogatz,Bernard} and identical chaotic elements \cite{Pecora2,Wu}.

In nature the interacting elements in complex networks are non-identical, in such situations complete synchronization is no longer possible, but a highly coherent state can be observed. 
Examples include collections of coupled maps \cite{Ott}, power grid networks \cite{Kurths}, superconducting Joseph junctions \cite{Strogatz1}, and brain networks \cite{Attention,Singer,Ep,Gorka}.
In these systems, coherence is characterized by the mean field controlling 
the behavior of the nodes. Of major importance is how the coherence properties of a general 
collection of non-identical nodes depends on the 
structural parameters of the network, on the coupling function, and on the 
node dynamics. Recent work has elucidated how the dynamics of the nodes can influence 
coherence in terms of mean field approximations for coupled maps \cite{Ott}, construction of a Lyapunov function \cite{Hasler}, and extending the Lyapunov exponents approach \cite{Bollt,Hill}. Moreover, control techniques have also been used \cite{Qiang,Cai}. Although, this problem has 
received increasing attention, the effect of network connectivity on the coherent systemic behavior still 
remains elusive. 

In this letter,  we uncover how the dynamical coherence of the network depends on the node dynamics, 
the coupling function,  
and the network connectivity. Our approach is fully analytical and holds  for a class of interaction functions whose Jacobian has positive real part spectrum.  We find that in purely random networks, coherence is controlled by the mean degree.
In locally connected networks, the mean degree no longer determines coherence, instead coherence is determined by an interplay between the network size and the degree of the nodes. If the mean degree scales properly with the network size, coherence emerges as the network grows. 
In these networks, even if coherence is absent, by adding random connections we can induce coherence.

The dynamics of a network of $n$ coupled elements with interaction akin to 
diffusion is described by 
%\vspace{-2pt}
\begin{equation}
\frac{d \bm{x}_i}{dt} = \bm{f}_i(\bm{x}_i) + 
\alpha \sum_{j} A_{ij} \bm{H}(\bm{x}_j- \bm{x}_i), 
\label{eq1}
\end{equation}
%\vspace{-1pt}
\noindent
where  $\bm{f}_i: \mathbb{R}^m \rightarrow \mathbb{R}^m$ is smooth and governs the dynamics of the isolated nodes.
$\bm{H} : \mathbb{R}^m \rightarrow \mathbb{R}^m$ is a smooth coupling function, $\alpha$ is the overall coupling strength, $A_{ij}=1$ if nodes $i$ and $j$ are connected 
and $A_{ij}=0$ otherwise. The degree, that is, the number of connections, of the $i$th node is given by 
$k_i = \sum_j A_{ij}$. In this model, coherence is
related to the Laplacian $\bm{L}$ where  $L_{ij} = \delta_{ij} k_i - A_{ij}$ and $\delta_{ij}$ 
is the Kronecker delta symbol.

We assume that the coupling function possesses  the following properties: 
$i)$ $\bm{H}(\bm{0}) = \bm{0}$, $ii)$ the Jacobian of the coupling function 
$D \bm{H}(\bm{0}) = \bm{\Gamma}$ has spectrum on the right part of the 
complex plane.
 Throughout the paper, $\beta>0$ denotes the smallest 
real part of the eigenvalues of $\bm{\Gamma}$.
The hypothesis on the spectrum of the Jacobian guarantees the
stability of the problem, and cannot be omitted, otherwise, instabilities could
appear due to an interplay between the dynamics of the individual nodes and the coupling function. 
 Here, the norm $\| \cdot\|$ 
denotes the Euclidean norm.

{\it Main result:} We consider $\bm{f}_i = \bm{f} + \bm{p}_i$, where $\| \bm{p}_i \| \le \delta$  
uniformly for all nodes \cite{Unif}.  We call $\delta$ the heterogeneity parameter. 
Our main finding is that, for large times, fluctuations of the trajectories 
are bounded by
%\vspace*{-1pt}
\begin{equation}
\| \bm{x}_i(t) - \bm{x}_j(t) \| \le \frac{K \delta}{ \alpha \beta \lambda_2 - \alpha_c}
\label{Main}
\end{equation}
%\vspace*{-1pt}
where $\delta$ measures the heterogeneity among the node dynamics, 
$K = K (\bm{\Gamma})$ is a constant, 
$\alpha$ 
is the interaction strength, 
$\alpha_c = \alpha_c(\bm{f},\bm{H})$ is positive if the isolated dynamics 
is chaotic, otherwise $\alpha_c=0$, and $\lambda_2 = \lambda_2(\bm{L})$ is the spectral gap, 
i.e., the second smallest eigenvalue of the Laplacian matrix $\bm{L}$. 
Roughly speaking, our assumptions will guarantee  that the coherence of the set of non-identical nodes
is determined by the synchronization properties of the system of identical nodes $\delta=0$.

{\it A network of identical nodes:} The situation 
of zero heterogeneity $\delta =0$ was studied in great detail in the past decades in terms of the master 
stability function \cite{Pecora2}. Here, we develop a theory based on dichotomies \cite{Martin}, 
and their persistence.  This allows us to develop a complete analytical treatment of the problem.
The fully synchronized state $\bm{x}_1 = \cdots = \bm{x}_n$ is invariant under the equation of motion
for all values of the coupling strength $\alpha$, and it is called {\em the synchronization manifold}.
To study the stability we expand the 
coupling function about the synchronization manifold, which yields
$
\bm{H}(\bm{x}_j - \bm{x}_i) = \bm{\Gamma} (\bm{x}_j - \bm{x}_i) + 
\bm{r}(\bm{x}_j - \bm{x}_i),
$
where $\bm{r}$ is a nonlinear Taylor remainder. 
We use the following convenient notation, denote 
$
\bm{X} = \mbox{col}( \bm{x}_1 , \cdots , \bm{x}_n ),
$
here col stands for the vector formed by stacking the column vectors
$\bm{x}_i$ into a single column vector, note that $\bm{X} \in \mathbb{R}^{n m}$. Similarly  
$\bm{F}(\bm{X}) = \mbox{col}( \bm{f}(\bm{x}_1) ,\cdots, \bm{f}( \bm{x}_n) ).$ Let $\otimes$ denote the tensor product. The dynamics of a network of  identical nodes can be represented in tensor form as 
$\bm{X}^{\prime} = \bm{F}(\bm{X}) - \alpha (\bm{L} \otimes \bm{\Gamma} ) \bm{X} + \bm{R}_H$, 
where $\bm{R}_H$ is the Taylor remainder of the coupling function.

The vector $\bm{1} =  (1,1,\cdots,1)/\sqrt{n}$, is an eigenvector of the Laplacian 
associated with the zero eigenvalue. Moreover, since the eigenvectors of $\bm{L}$ are orthogonal, we
consider the following decomposition
$\bm{X} = \bm{1} \otimes \bm{s} + \bm{\xi}
$, where $\bm{\xi}$ does not lie in the span of $\bm{1} \otimes \bm{s}$.  Note that if 
$\bm{\xi}$ is zero, then the system is fully synchronized. The variational equation governing the $\bm{\xi}$ reads as $\bm{\xi}^{\prime} = \bm{K}(t) \bm{\xi}$, with $\bm{K}(t) =   
\bm{I}_n \otimes D \bm{f}(\bm{s}(t)) - \alpha \bm{L} \otimes \bm{\Gamma}$, where we neglected the nonlinear terms. The unique
solution of this  equation can be  represented in terms of the evolution
operator $ \bm{\xi}(t) = \bm{T}(t,s) \bm{\xi}(s) $. 
{ We postpone technical manipulations \cite{Longer_version}} and present 
the main result concerning the contraction properties of  the evolution operator
$
\|  \bm{T}(t,s) \| \le  K \exp\{ - [\alpha \beta \lambda_2 - \alpha_c] (t-s)\}
$
for any $t\ge s$, where $K = K(\bm{\Gamma})$ is a constant independent of the network, $\alpha$ is the coupling strength,
$\beta>0$ is the smallest eigenvalue of $\bm{\Gamma}$, and   
$\lambda_2$ is the spectral gap of the Laplacian. Here,  $\alpha_c =
\alpha_c(\bm{f},\bm{H})$, and is positive if the node dynamics 
$\bm{f}$ has positive Lyapunov exponents. Note that since the contraction of the evolution operator is uniform the nonlinear remainders will not 
effect the stability of the transient towards synchronization.
This finding implies that starting at a time $s$ with nearby initial conditions we obtain 
$
 \| \bm{x}_j(t) - \bm{x}_i(t) \| \le C e^{- (\alpha \beta \lambda_2 - \alpha_c)(t-s)}.
$
for all $t\ge s$. Therefore, the characteristic decay time is $1/[\alpha \beta \lambda_2 - \alpha_c]$. 
Next, we show that the characteristic time controls the coherence of the heterogeneous network.

{\it Effect of the  Heterogeneity:} We consider Eq. (\ref{eq1}) 
with node dependent maps $\bm{f}_i$. 
Again, we represent the equations in the tensor representation, and perform the same decomposition as before 
$\bm{X} = \bm{1} \otimes \bm{s} + \bm{\xi}$. The equation of motion can now be written as $\bm{X}^{\prime} = \bm{F}(\bm{X}) - \alpha (\bm{L} \otimes \bm{\Gamma} ) \bm{X} + \bm{P}(\bm{X}) + \bm{R}_H(\bm{\xi})$, 
$\bm{P} = $ col $(\bm{p}_1, \cdots , \bm{p}_n)$. 
We project the equation onto the synchronization manifold and to the 
orthogonal complement. The first projection gives us an equation for $\bm{1} \otimes \bm{s}$ and the second an equation for perturbations $\bm{\xi}$. After some manipulations, the equation for the perturbations reads
 $ \bm{\xi}^{\prime} = \bm{K}(t) \bm{\xi} + 
\bm{G}(\bm{X})$, where  $\bm{G}(\bm{X})$ is the  projection of $\bm{P}$, $\bm{R}_H$ and $\bm{R}_F$ onto the orthogonal complement of the synchronization manifold. Here, $\bm{R}_F$ is the Taylor remainder of the vector field $\bm{F}$ about the synchronization manifold. By the method of variation of  parameters the solution becomes
$\bm{\xi}(t) = \bm{T}(t,0)\bm{\xi}(0) 
+ \int_0^t \bm{T}(t,u) \bm{G} (\bm{X}(u))du.$ Then, 
using the bounds on the norm of the evolution operator together with 
the triangle inequality for large times we obtain Eq. (\ref{Main}) follows. 

We use Eq. (\ref{Main}) to study the effect of network connectivity. To this end, 
we measure the macroscopic coherence in the network by introducing the quantity
%\vspace*{-3pt}
{
\begin{equation}
E(t) = 1 -  \frac{\sum_{i,j} \|   \bm{x}_i(t) - \bm{x}_j(t) \|}{ n(n-1) V },
\label{E}
\end{equation}
%\vspace*{-1pt}
\noindent
which quantifies coherence as a measure of the distance between the trajectories of the nodes per link. In Eq. (\ref{E}), $V$ 
is a normalization factor $V = \max\| \bm{x}_i - \bm{x}_j \|$ for $\alpha = 0$ (when there is no interaction).  
If the nodes are uncorrelated then $E \rightarrow 0$, the more coherent the system is $\| \bm{x}_i - \bm{x}_j \|$ approaches zero, 
and $E(t)  \rightarrow 1$. } Hence, when the mean field dominates the dynamics of the individual nodes, we obtain
$
|1 - E(t) |  \propto  \delta/ (\alpha \beta \lambda_2 - \alpha_c) .
$ 
The denominator 
must be positive, so if $\alpha_c > 0 $ and $\lambda_2 \rightarrow 0$ as $n \rightarrow \infty$, 
we can loose coherence at a finite network size.  To illustrate these findings, we show that in a $2k$ nearest 
neighbor network there is a {\it critical number of neighbors} as a function 
of the network size $n$ for the transition to coherence. 

For concreteness, we explore these findings using the Lorenz equation exhibiting
a chaotic dynamics \cite {Viana} to represent the node dynamics. 
Using the notation 
$
\bm{x}_i = 
(x_i, y_i, z_i)^*$, where $^*$ denotes the transpose, the vector field reads
$
\bm{f}(\bm{x}) = (\sigma ( y -x ), x ( r  - z) - y, -b z + xy 
)^*,
$ we choose the classical parameter values $\sigma=10, r = 28, b = 8/3$.
We consider the non-identical behavior as a mismatch in the parameter $\sigma$. 
Hence, each Lorenz system has $\sigma_i = \sigma + \zeta_i$, where
$\zeta_i$ is a random number picked independently according to a uniform distribution with 
support $[-\varepsilon,\varepsilon]$, yielding 
$\bm{p}_i(\bm{x}_i) = (\zeta_i (y_i - x_i),0,0)^*$. Hence, 
$\| \bm{p}_i(\bm{x}_i) \| \le  M \varepsilon$, where $M$ is such that 
$|x - y|< M$, for the Lorenz $M \approx 40$. Note that with this choice the heterogeneity is $\delta = 2 M \varepsilon$.   For simplicity we choose $\bm{H}(\bm{x})=\bm{x}$. We fix the 
coupling strength $\alpha = 10$.
The trajectories of the Lorenz accumulate in a neighborhood of a chaotic attractor, and hence, $\alpha_c >0$ \cite{Alphac}. 
For our numerical simulations, we used the 4th order Runge-Kutta integration scheme with integration step $10^{-4}$.

{\it Locally Connected Networks:} Consider a network of $n$ nodes, in which each
node is coupled to its $2k$ nearest neighbors. 
The network Laplacian can be diagonalized and the spectral gap explicitly obtained
$\lambda_2 = 2k+1 - {\sin[(2k+1)\pi / n]  / \sin(\pi/n)}$ \cite{Kurths}.  We analyze  $k$ as a function of the network 
size $n$ as $k =  \lceil n^{\gamma}/2 \rceil$ -- 
where $\lceil x \rceil$ is the largest integer that approximates $x$. 
It follows that there exists a {\it critical number of neighbors}
needed for the network to self organize towards  coherence, that is,  
there is a critical $\gamma_c$ such that for 
$\gamma> \gamma_c$ there is a transition to self organization and coherence is enhanced as the
network size increases: the heterogeneity is suppressed and the mean 
field dominates the dynamics. In contrast, for 
$\gamma< \gamma_c$ coherence is absent. 
The value of $\gamma_c$ for onset of coherence can be predicted by analyzing the zero of the 
denominator of Eq. (\ref{Main}).  For 
$k \ll n$ we obtain, up to the leading order in $n$, the following equation for the critical value of $\gamma_c$:
%\vspace*{-2pt}
\begin{equation}
\gamma_c = \frac{1}{3} \left[ 2 - \frac{\log\left( \frac{\pi^2 \alpha \beta}
{6 \alpha_c}\right)}{\log n}  \right]. 
\label{gc}
\end{equation}
%\vspace*{-2pt}

\begin{figure}[t]
\centering
\includegraphics[scale=0.45]{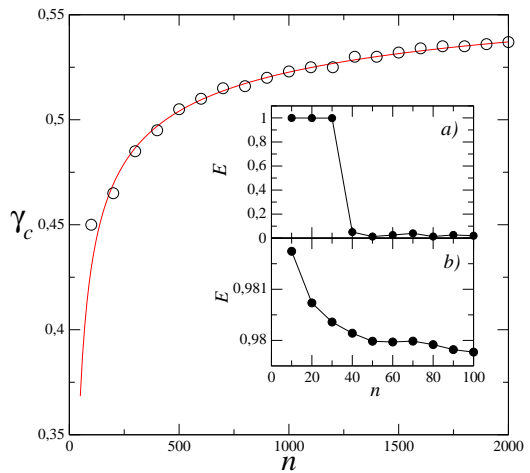}
\caption{$\gamma_c$ as a function of the network size $n$ for $k = \lceil n^{\gamma}/2 \rceil$.  
We fixed $\alpha = 10$ and $\varepsilon = 0.2$, and recall $\beta = 1$. The open circles  
represent the numerical determination of $\gamma_c$, and  the solid line is the theoretical 
prediction, obtained  solving $\alpha \beta \lambda_2 - \alpha_c = 0$  
for $\gamma$.  The inset $a)$ 
shows the critical behavior of $E(t)$ for 
$\gamma=0.3$ as a function of the network size. Here at the critical network size $n=41$ coherence 
is lost, agreeing with the theoretical predictions.  
Inset $b)$ we present the same numerical simulations of the Lotka-Volterra model. As we predict 
no abrupt transition is observed once $\alpha_c = 0$.}
\label{Fig1}
\end{figure}

We check these predictions against the numerical simulations of the Lorenz dynamics.  
We consider $\varepsilon=0.2$, and for each fixed system size $n$ 
considering $k =  \lceil n^{\gamma}/2 \rceil$,  we vary $\gamma$ and measure 
$E(t)$, see Eq. (\ref{E}).  The value of $\gamma_c$ is determined by observing the 
behavior of $E$.  { Typically, $E \approx 0$ before the transition to coherence, 
and after the transition $E \approx 0.97$}. Since, the transition from an incoherent to a 
coherent state is sharp, we can easily detect the value of $\gamma_c$. This numerical 
determination of $\gamma_c$ is presented as open circles in Fig. \ref{Fig1}, against the 
theoretical prediction presented as a solid line. 
Likewise, for a fixed $\gamma$ we can vary the system size and determine the transition 
in $E$. Again, we find the 
critical network size as a function of $\gamma$. 
In the inset a) of  Fig. \ref{Fig1}, we exhibit a case where we fixed $\gamma = 0.3$ and 
varied the system size. A sharp transition towards loss of coherence can be observed for $n = 41$.  

%\vspace*{-5pt}
\noindent
%\vspace*{-10pt}

If the isolated dynamics has $\alpha_c=0$, then there is no abrupt transition towards coherence. Either
an enhancement of coherence for $\gamma > \gamma_c$, or a deterioration for $\gamma < \gamma_c$.
An example of this situation is observed in the standard Lotka-Volterra model. 
The state vector $\bm{x}$ of the model is two dimensional, and the vector field reads $\bm{f}(\bm{x}) 
= (x[a - by],  y [-d + cx])^*$. This system has a constant of motion, 
which means that $\alpha_c = 0$. 
For simplicity we consider all parameters equal to $1$ and mismatches in the 
parameter $a$ as $a_i = a + \zeta_i$, which $\zeta_i$ as before.   In our simulations on $2k$ 
nearest neighbor network, we fixed $\varepsilon=0.2$, $\gamma=0.3$ and vary the network size $n$. 
We present the results in the inset $b)$ of Fig. \ref{Fig1}. We observed no abrupt transition to loss of 
coherence, in agreement with our predictions. 

\begin{figure}[h]
  \centerline{\hbox{\psfig{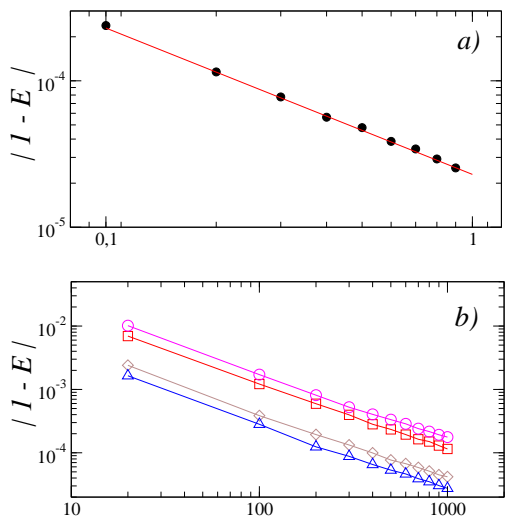}}}
\caption{ { Randomness enhance coherence. In $a)$ 
Log-Log plot of coherence $|1-E|$ versus the fraction of randomly added links $s$. For fixed $\alpha = 10$. 
starting from a $2k$ nearest neighbor network of size $n=1000$ and $\gamma=0.3$ no coherence is observed.  
We then add a fraction of 
$s$ random links. This induces coherence in the network according to $|1 - E | \propto s^{-1}$, as predicted theoretically.
In $b)$ 
Log-Log plot of coherence $|1-E|$ versus the mean degree $m$. For  $n=3000$, we simulate the 
Lorenz dynamics on Erd{\"o}s-R{\'e}nyi with $\varepsilon = 1 $ ($\square$), Power law networks with 
$\theta = 3$ and $\varepsilon = 1$ ($\circ$).  Then with $\varepsilon = 0.2$ for $\theta = 2.7$ ($\diamond$) and 
$\theta = 4$ ($\bigtriangleup$). The scaling towards coherence $|1 - E| \propto \delta / m^{-1}$ agrees with 
the theoretical prediction.}}
\label{Fig2}
\end{figure}

{\it Small World graphs: Enhancing Collective Motion.} 
Starting from a nearest neighbor network where no coherence is observed, we can enhance coherence 
by adding in a small fraction of random links. We add $sn$ edges picked at random from the remaining 
unconnected pairs, so that the average number 
of shortcuts per node is $s$. This new network is called {\it small world}. A perturbation theory allows
to estimate the expected values of the Laplacian spectral gap. Computing  the eigenvalues of the 
Laplacian perturbatively
reveals that for ${1/3} < \gamma < 1$, also 
considering $n \gg 1$,  we obtain that $\lambda_2 = 2s + O(n^{\gamma - 1}) $ is 
the expected value of the eigenvalue.  As before, the denominator 
of Eq. (\ref{Main}) must be positive, and this provides a critical number of shortcuts $s_c$ that must be added 
to obtain coherence. If $s$ is larger then $s_c$ we undergo a transition to onset of coherence and, up to high order 
corrections in the system size, where the coherent measure is given by
$
|1- E(t)| \propto \delta/  s.
$
We check this prediction against numerical simulations using the Lorenz dynamics to model 
the nodes. Starting from the nearest neighbor network of size $n=1000$ and $\gamma = 0.3$ (no 
coherence is observed), we add  a fraction of $s$  random edges. We then vary $s$ and 
measure the coherence. The result can be observed in the inset Fig. \ref{Fig2} $a)$.
The theoretical prediction is in excellent agreement  with our simulations of the Lorenz dynamics. 
Thus, by adding a small fraction of random connections  we induce coherence. Here, the enhancement
of coherence is proportional to the fraction of random shortcuts $s$.

{\it Random Networks:} As we discussed in the previous paragraph random structures can 
enhance coherence. In purely random networks, the coherence is proportional to the mean degree.  
We use a random graph model $G(\bm{w})$ for a sequence of expected degrees 
${\bm w} = (w_1 , w_2 ,  \cdots, w_n )$. 
Each element of the adjacency $A_{ij}$'s is an independent Bernoulli variable, taking 
value $1$ with success probability 
$p_{ij} = w_i w_j \rho,$ 
where $\rho = 1/ \sum_{i=1}^n w_i.$ 
The sequence must satisfy the condition $w_1^2 \le \rho$ to assure that $p_{ij} \le 1$.
Under these constructions $w_i$ is the expected value of $k_i$.  
The mean degree $m = 1 /( n \rho)$ determines 
the spectral gap $\lambda_2$. 
The Erd{\"o}s-R{\'e}nyi random graphs correspond to 
the constant case $w_i = p n$. If $p$ is constant then { the expected value of  $\lambda_2$ is concentrated at $m$.}
The power law graphs correspond to the case, $w_i \propto i^{-1/(\theta-1)}$, with $\theta \ge 2$. 
  See Ref. \cite{CombChung} for details 
on this choice of $w_i's$. 
The parameter $\theta$ characterizes the degree distribution, that is, the probability $P(k)$ to find a 
degree between $k $ and $k + \Delta k$, 
behaves as a power law $P(k) \propto k^{-\theta }$.
If the network is large, { the expected value of the spectral gap 
$\lambda_2$ is concentrated at $m(1 - 1/(\theta -1))$}, see Ref. \cite{Wu}. 
In these cases, for large mean degrees $m$, we obtain the scaling $|1 - E| \propto \delta/m$.
%\vspace*{-10pt}
%\vspace*{-10pt}
\noindent
We constructed these random networks with size $n=3000$ and studied numerically the coherence 
properties as a function of the mean degree $m$ and heterogeneity $\delta = 2M \varepsilon$.   Our numerical
 simulations using the Lorenz dynamics yield $ E = 1 - O(\delta / m)$, in excellent agreement with our 
 predictions, see Fig. \ref{Fig2} b).

In summary, we have uncovered the dependence of network coherence on 
the dynamics of the nodes, the network connectivity and the coupling function. 
In random networks, dynamical coherence is enhanced with the increase of the mean degree. 
{ These networks exhibit high connectivity. 
In regular networks the mean degree no longer controls the emergence and enhancement of coherence, rather 
we encounter a critical behavior: if the mean degree scales properly with the 
system size coherence emerges. We were able to determine such critical behavior analytically. 
In our numerical illustrations, we chose the non-identical part $\bm{p}_i$ as a mismatch component. 
Our approach is general and $\bm{p}_i$ can be an essentially different system, or a noise driven component. 
In the later case, our results predict a noise suppression due to network effects.  }
Formula (\ref{Main}) explains how the connectivity can enhance coherence, which can be useful for many 
applied areas where coherence plays a fundamental role such 
power grid networks and neuroscience. 

We are in debt with A. Pikovsky, R. Vilela, and J. Eldering for illuminating conversations.
This work was financially supported by Leverhulme Trust Grant No. RPG-279, 
CNPq, TUBITAK Grant No. 111T677. 
%UT is a member of the Science Academy, Turkey.

%\vspace{-20pt}


\begin{thebibliography}{99}
\bibitem{Unif} If $\bm{p}=\bm{p}(t)$ we consider the 
following norm $\| \bm{p} \| = \sup_{t \ge 0} \| \bm{p}(t)\|$. We also assume for simplicity that 
the dynamics of each perturbed node lies within a compact set.
\bibitem{Alphac} 
Our  analysis provides an estimate for $\alpha_c$, this bounds is not sharp. Hence, we use 
the bounds from the theory of the Lyapunov exponents. However, we could loose coherence due to bubbling.  
Statistically, these scenarios occur with small probability. Therefore, $\alpha_c \approx 0.9$, and average our 
numerical experiments over many trials.
\end{thebibliography}
\end{document}